\documentclass[12pt]{article}
\usepackage{amsmath,amssymb}
\usepackage{cancel}
\usepackage{bm}
\usepackage{braket}
\usepackage{cite}
\usepackage[pdftex]{graphicx}

\makeatletter

\@addtoreset{equation}{section}
\makeatother

\newcommand{\fs}{\hspace*{-4pt}}

\begin{document}

\title{
\baselineskip 14pt
\hfill \hbox{\normalsize EPHOU-20-003}\\ 
\hfill \hbox{\normalsize WU-HEP-20-03} \\
\hfill \hbox{\normalsize KUNS-2788}
\vskip 1.7cm
\bf
Loop Fayet-Iliopoulos terms in $T^2/Z_2$ models: \\ 
instability and  moduli stabilization
\vskip 0.5cm
}
\author{
\centerline{Hiroyuki~Abe$^{1}$, \
Tatsuo~Kobayashi$^{2}$, \
Shohei~Uemura$^{3}$ and \
Junji~Yamamoto$^{4}$}
\\*[20pt]
\\
{\it \normalsize 
\centerline{${}^{1}$Department of Physics, Waseda University, Tokyo 169-8555, Japan}}
\\
{\it \normalsize 
\centerline{${}^{2}$Department of Physics, Hokkaido University, 
Sapporo 060-0810, Japan}}
\\
{\it \normalsize 
\centerline{${}^{3}$CORE of STEM, Nara Women's University, Nara 630-8506, Japan}}
\\
{\it \normalsize 
\centerline{${}^{4}$Department of Physics, Kyoto University, 
Kyoto 606-8502, Japan}}
\\*[50pt]}

\date{
\centerline{\small \bf Abstract}
\begin{minipage}{0.9\linewidth}
\medskip 
\medskip 
\small
We study Fayet-Iliopoulos (FI) terms of six-dimensional supersymmetric Abelian gauge theory compactified on a $T^2/Z_2$ orbifold.
Such orbifold compactifications can lead to localized FI-terms and instability of bulk zero modes.
We study 1-loop correction to  FI-terms in  more general geometry than the previous works.
We find induced FI-terms depend on the complex structure of the compact space.
We also find the complex structure of the torus can be stabilized at a specific value corresponding to  a self-consistent supersymmetric minimum of the potential by such 1-loop corrections, which is applicable to the modulus stabilization.
\end{minipage}
}

\newpage

\begin{titlepage}
\maketitle
\thispagestyle{empty}
\clearpage
\end{titlepage}

\section{Introduction}

Effective theory of superstring includes various dimensional objects, i.e., branes.
Branes are important components for particle phenomenology.
Branes can break the supersymmetry (SUSY) and realize the chiral spectrum \cite{Blumenhagen:2000wh, Bachas:1995ik, Berkooz:1996km, Aldazabal:2000dg}.
They can be a source of generations of matter fields, and flavor structure \cite{Ibanez:2001nd, Abe:2008sx}.
Anti-branes can induce the positive cosmological constant \cite{Kachru:2003aw}.
Such a brane mode behaves as a localized mode in effective theory.
Therefore it is important to investigate interactions between bulk fields and localized operators \cite{ArkaniHamed:2001tb, Ishida:2017avx}.

The Fayet-Iliopoulos term (FI-term) in supersymmetric Abelian gauge theory was introduced as a source of spontaneous
SUSY breaking at first \cite{Fayet:1974jb}.
Later it was shown that FI-term is not only a source of the SUSY breaking,
but has vast implications for theoretical particle physics.
The FI-term is prohibited by local SUSY unless the gauge group is related to $U(1)_R$ \cite{Freedman:1976uk, Barbieri:1982ac, Binetruy:2004hh}
or associated with non-linear terms \cite{Cribiori:2017laj}.
Especially in higher dimensional supersymmetric theory,
it is related to anomaly \cite{Barbieri:2002ic}, and introduces instability of bulk superfields \cite{GrootNibbelink:2002qp, GrootNibbelink:2002wv}.(See also \cite{Abe:2002ps}.)

Even in the higher dimensional theory,
the bulk FI-term is prohibited by local SUSY,
but the FI-term localized at special points, i.e., orbifold fixed points can appear \cite{Abe:2004yk}.
Such a FI-term is called localized FI-term.
The localized FI-term is induced by quantum corrections in orbifold compactification
even if the FI-term is set to zero at the tree level \cite{Ghilencea:2001bw}.
This is formally calculated by infinite sum of all KK-modes of fields
which have charges of the corresponding $U(1)$.
In the trivial background without the localized FI-term, mode expansion of bulk fields is given by plane waves.
Their infinite sum converges to the Dirac delta function.
Hence the localized FI-term is induced.
Since it is localized, the FI-term induces a local potential for bulk fields.
To cancel the FI-term, the vacuum expectation values (VEVs) of auxiliary fields must also be localized.
It affects the wave function profiles of the bulk fields.
For the model of five-dimensional Abelian gauge theory compactified on $S^1/Z_2$,
the localized FI-term induces localization of bulk zero modes at the fixed points, 
and rejects wave functions of all the massive modes from the fixed points \cite{GrootNibbelink:2002qp, GrootNibbelink:2002wv}.
Similar results are obtained also for six-dimensional SUSY theory compactified on $T^2/Z_2$ orbifold \cite{Lee:2003mc}.
Thus it is a quite general consequence for higher dimensional SUSY theory
compactified on orbifolds.

If the value of the localized FI-term is not zero, 
VEVs of the auxiliary fields are shifted.
The massive modes can not penetrate to the fixed points in this 1-loop corrected vacuum.
Hence 1-loop corrections to the FI-terms are  only due to the zero mode.
The zero mode is localized at the fixed points, and reproduces the localized FI-term,
but it is not the same as that of the infinite sum of the plane waves.
Bulk contribution is not canceled by  brane mode contributions in general,
and the FI-term receives further corrections.
Thus this background is unstable.
In our previous work we investigated this instability for the $S^1/Z_2$ compactification model \cite{Abe:2019nkv}.
In the present paper we investigate instability for $T^2/Z_2$ compactification.
Toroidal compactification is a more realistic compactification for phenomenology;
it has a concrete stringy origin \cite{Blumenhagen:2000wh}.
It also can realize the chiral spectrum of the Standard Model (SM). (See e.g. Refs.~\cite{Abe:2008fi, Abe:2012fj}.)
The localized FI-term on toroidal orbifold may affect the flavor structure of the SM \cite{Abe:2018ylo}.
As well as $S^1/Z_2$ compactification, 
loop correction of the FI-term can lead to the instability of 1-loop corrected vacuum.
We find that the instability is related to the complex structure of the torus.
There are some applications for moduli stabilization and extra dimensional models.

This paper is organized as follows.
In section 2, we examine the localized FI-term and zero mode of bulk scalar field in six-dimensional SUSY gauge theory compactified on $T^2/Z_2$ orbifold, whose geometry is described by an arbitrary value of the complex structure modulus $\tau\, (\in \mathbb{C})$.
The localized FI-term is induced by quantum corrections, 
and it leads to nonzero VEVs of auxiliary fields.
It affects equations of motion for bulk fields and their wave function profiles.
In Section 3, we focus on an untilted torus, i.e., a torus whose complex structure is pure imaginary,
and recalculate the 1-loop corrections to the FI-term in the SUSY vacuum which has nonzero VEV of the auxiliary field.
We see that 1-loop corrections can cause the instability of the SUSY vacuum.
In Section 4, we extend the consequences in section 3 to the torus that has an arbitrary value of  $\tau$.
We find these quantum corrections depend on $\tau$.
We show the complex structure modulus must take a specific value for the cancellation between loop corrections from bulk and brane modes.
In other words, modulus stabilization of the complex structure is realized.
Section 5 is devoted to our conclusion.
In Appendix \ref{app:local}, we study the validity of our evaluation of the FI terms. We also confirm the localization of the wave function of bulk zero mode by use of an explicit regularization of the Dirac delta function.
In Appendix \ref{app:C}, we show the modular transformation of elliptic theta functions.

\section{Localized FI-terms on $T^2/Z_2$ model}

In this section, we evaluate the localized FI-term induced by quantum corrections in the $T^2/Z_2$ orbifold.
We take the following strategy.
First we consider 1-loop FI-term induced by tree level wave functions.
Then we investigate mode expansion of the bulk fields in the 1-loop corrected background including a singular configuration of the gauge field.
Finally we recalculate the quantum correction of the FI-term induced by the 1-loop corrected wave functions, and search a consistent configuration.

Before describing the multiplets that are contained in $T^2/Z_2$
models, we describe the torus $T^2$ and orbifold action of $Z_2$.
We define the orthogonal coordinates of $T^2$ as $x_5, \, x_6$, and we denote the two-dimensional metric by $g_{ij}$:
\begin{align}
	g_{ij} =\begin{pmatrix} 1& 0\\ 0&1 \\ \end{pmatrix} \,\,\,\,\, (i,j = 5,\,6).
\end{align}
The coordinates $(x_5,\, x_6)$ satisfy the following periodic boundary conditions:
\begin{align}
\begin{cases}
(x_5, x_6) \sim (x_5 +2\pi R, x_6),
\nonumber\\
(x_5, x_6) \sim (x_5 +2\pi R \, \text{Re}\, \tau,  x_6+2\pi R \, \text{Im}\, \tau).
\nonumber
\end{cases}
\end{align}
where we introduced a complex structure $\tau$, which takes an arbitrary value in the upper half plane $\mathbb{H}$.
We define the $Z_2$ orbifold action as
\begin{align}
	Z_2 : (x_5, \, x_6) \rightarrow (-x_5,\, -x_6),
\end{align}
and there are four fixed points: $(0,0), (\pi R,0), (\pi R\ \text{Re} \tau,\pi R\ \text{Im}\, \tau)$ and $(\pi R(1+\text{Re} \tau),\pi R \ \text{Im}\, \tau)$.
Hereafter these fixed points are denoted by $z_1, z_2, z_3$ and $z_4$, respectively.

We introduce non-orthogonal coordinates $(x'_5,\, x'_6 )$ which are along the lattice vectors of the torus.
In these non-orthogonal coordinates, the two periodic boundary conditions can be represented as
\begin{align}
\begin{cases}
	(x'_5, x'_6) \sim (x'_5+2\pi R, x'_6), \\
	(x'_5, x'_6) \sim (x'_5, x'_6 +2\pi R).
\end{cases}
\end{align}
We also define complex coordinates $(z, \, \bar{z})$ as $Rz \equiv x'_5 +\tau x'_6$ and $R\bar{z} \equiv x'_5 + \bar{\tau} x'_6$.
From now on, we use the notation of indices as $M,N\in\{0,1,2,3,5,6\}$, $\mu,\nu \in\{0,1,2,3 \} $, and $i,j,m,n\, \in \{5,6\}$.
We also use the indices with prime, $M',N',i',j'$ to represent the non-orthogonal coordinates $(x'_5, x'_6)$.
We summarize the relations of the coordinates and the metrics in Table \ref{tab:coordinate}.

\begin{table}
\centering
\begin{tabular}{|c||c|c|} \hline
 & \text{Non-orthogonal}  & \text{Complex} \\ \hline \hline
\text{coordinates} &  $x'_5 = x_5 -\frac{\text{Re}\tau}{\text{Im}\, \tau} x_6$ & $Rz=x'_5+\tau x'_6$ \\
 &  $x'_6=\frac{1}{\text{Im}\, \tau}x_6$ & $R\bar{z}=x'_5+\bar{\tau} x'_6$ \\ \hline
\text{boundary } &   $x'_5\sim x'_5+2\pi R$ & $z\sim z+ 2\pi$ \\
\text{ conditions} &  $x'_6\sim x'_6 +2\pi R$ &  $z\sim z+2\pi \tau$ \\ \hline
\text{metric}&
$g_{i'j'} =\begin{pmatrix} 1 & \text{Re}\tau \\ \text{Re}\tau &|\tau|^2 \\ \end{pmatrix}$ &  $g_{mn} = \frac{1}{R^2}\begin{pmatrix} 0& 2 \\ 2 &0 \\ \end{pmatrix}$ \\ \hline
\end{tabular}
\caption{Coordinates and metrics on the torus}
\label{tab:coordinate}
\end{table}

We consider six-dimensional SUSY Abelian gauge theory defined below the cutoff scale $\Lambda$.
Such a theory is described by four-dimensional $\mathcal{N} =2$ supermultiplets: Abelian vector multiplet and hypermultiplets.
In addition to the $\mathcal{N} =2$ multiplets, we can introduce brane modes at the fixed points.
The brane modes preserve $\mathcal{N} =1$ SUSY and we assume that they consist of only chiral multiplets; there are no extra gauge fields at the fixed points.
We introduce brane mode $\Phi_I =(\phi_I, \psi_I)$ at each fixed point $z_I$.
The multiplets are summarized as follows: 
\begin{align}
	\bullet \,&\text{bulk mode:}
	\begin{cases}
	\mathcal{N}=2\text{ Abelian vector multiplet }\\ \,\,=
\text{ \{gauge field }A_M,\text{ gaugino }\Omega,\text{ auxiliary field }\vec{D}\} \\
	\text{hypermultiplet }= \text{ \{real scalars $A_i$,\, hyperino $\zeta$\}}
	\end{cases} \nonumber \\ \nonumber 
	\bullet \, &\text{brane mode: }
	\text{chiral multiplet $=$ \{complex scalar $\phi_I$,\, Weyl fermion $\psi_I$ \}} .
\end{align}
We should pay attention to the auxiliary fields in $\mathcal{N}=2$ Abelian vector multiplet.
It is decomposed into an $\mathcal{N} =1$ vector multiplet and a single chiral multiplet.
The auxiliary field $D$ of the $\mathcal{N} =1$ vector multiplet is given by a linear combination of a part of the auxiliary field $\vec{D}$ and the field strength $F_{56}$.
We choose $D= - D_3+F_{56}$ in this paper.
The $Z_2$ orbifold action is defined to preserve this four-dimensional $\mathcal{N} =1$ structure, e.g., the parity assignment to $D_3$ is even and those to other two auxiliary fields $D_1$ and $D_2$ are odd.
We also introduce two complex scalar fields $\phi_+$ and $\phi_-$, which are linear combinations of the real scalars of the hypermultiplet.
$\phi_+$ is parity even and $\phi_-$ is parity odd.\footnote{For precise calculation, see \cite{Lee:2003mc}.}

The bosonic Lagrangian 
is written as follows:
\begin{align}
	\mathcal{L}=& -\frac{1}{4} F_{MN}F^{MN} + i \bar{\Omega} \Gamma^M \partial_M \Omega +\frac{1}{2}\vec{D}^2 
	+\sum_{\pm} (\mathcal{D}_M \phi_\pm^\dagger \mathcal{D}^M \phi_\pm \mp g\phi_\pm^\dagger q \phi_\pm D_3)+\cdots \notag \\
	&+\sum_{I=1}^{4} \delta (x_5-x_5^I)\,\delta(x_6-x_6^I) \Big[ \mathcal{D}_\mu \phi_I^\dagger \mathcal{D}^\mu \phi_I +g\phi_I^\dagger q_I \phi_I (-D_3 + F_{56}) + \cdots \Big],
\label{Action_tilted}
\end{align}
where 
\begin{align}
\mathcal{D}_M \phi_\pm=\partial_M \phi_\pm \pm igq \phi_\pm A_M . \notag
\end{align}
The quantities $q$ and $q_I$ are charges of the hypermultiplet and the brane modes respectively.
$g$ is the gauge coupling constant.
Four-dimensional effective potential is represented as follows:
\begin{align}
		V_{4d} = \int dx_5 dx_6\Big[& 2g^2 |\phi_+^T q \phi_-|^2+\sum_{\pm} (\mathcal{D}_5 \phi_\pm +i \mathcal{D}_6 \phi_\pm)^\dagger (\mathcal{D}_5 \phi_\pm +i \mathcal{D}_6 \phi_\pm) \nonumber \\
		&+\frac{1}{2} \big( F_{56} -\xi -g(\phi_+^\dagger q \phi_+-\phi_-^\dagger q\phi_-) -g\sum_I\phi_I^\dagger q_I \phi_I \delta^{(2)}(x_5-x_5^I, x_6-x_6^I) \big)^2  \nonumber \\
		&-\frac{1}{2} \big( D_3 -\xi -g(\phi_+^\dagger q \phi_+-\phi_-^\dagger q\phi_-) -g\sum_I\phi_I^\dagger q_I \phi_I \delta^{(2)}(x_5-x_5^I, x_6-x_6^I) \big)^2 \nonumber \\
		& -\frac{1}{2} (D_1 +g\phi_+^T q\phi_-+g\phi_-^\dagger q\phi_+^\ast)^2-\frac{1}{2}(D_2 +ig\phi_+^T q\phi_--ig\phi_-^\dagger q\phi_+^\ast)^2 \Big], \label{4DEP}
\end{align}
where we include the contributions of FI-term $\mathcal{L}_{\text{FI}} = \xi (-D_3+F_{56})$.

From (\ref{4DEP}), the SUSY conditions are written by
\begin{align}
&D_3= F_{56} = \xi +g(\phi_+^\dagger q \phi_+-\phi_-^\dagger q\phi_-) +g\sum_I\phi_I^\dagger q_I \phi_I \delta^{(2)}(x_5-x_5^I, x_6-x_6^I),
\\
&~~~~~~~~~~~~~~~~~~~~~~~~\phi_+^T q \phi_- = 0,
~~~~~~\mathcal{D}_5 \phi_\pm +i \mathcal{D}_6 \phi_\pm=0.
\end{align}
We study the situation where the $U(1)$ is unbroken, i.e., $\braket{\phi_\pm}=\braket{\phi_I}=0$.
The SUSY 
solution is as follows:
\begin{align}
\braket{F_{56}} = \xi(x_5,\, x_6). \label{SUSY_Cond}
\end{align}
We also obtain the equation of motion (EOM) for the scalar fields $\phi_\pm$ in terms of the compact directions:
\begin{align}
	& \text{(zero mode) : }(\mathcal{D}_5+i\mathcal{D}_6)\phi_\pm=0, \label{zero_mode_EOM} \\
	& \text{(massive mode) : }(-\mathcal{D}_5+i\mathcal{D}_6)(\mathcal{D}_5+i\mathcal{D}_6)\phi_\pm = \lambda \phi_\pm. \label{massive_mode_EOM}
\end{align}

The SUSY solution and the zero mode equation in the non-orthogonal coordinates are represented simply as follows:
\begin{align}
	\text{(SUSY condition) : }&\braket{F_{5'6'}} = |\text{Im}\, \tau|\xi (x'_5, \,x'_6), \label{SUSY_Cond'} \\
	 \text{(zero mode EOM) : }&(\tau \mathcal{D}_{5'}-\mathcal{D}_{6'})\phi_\pm (x'_5, \,x'_6)=0. \label{zero_mode_EOM'}
\end{align}
By evaluating (\ref{SUSY_Cond'}) and (\ref{zero_mode_EOM'}), we will confirm that, 
if the localized FI-term has a nonzero value,
the zero mode of the bulk field is localized at the fixed points $z=z_I$, that is similar to \cite{Lee:2003mc}.

\subsection{KK-modes and 1-loop FI-term when $\xi=0$}\label{Sec:No Background}

\begin{figure}[thbp]
\centering
\includegraphics[scale=0.25]{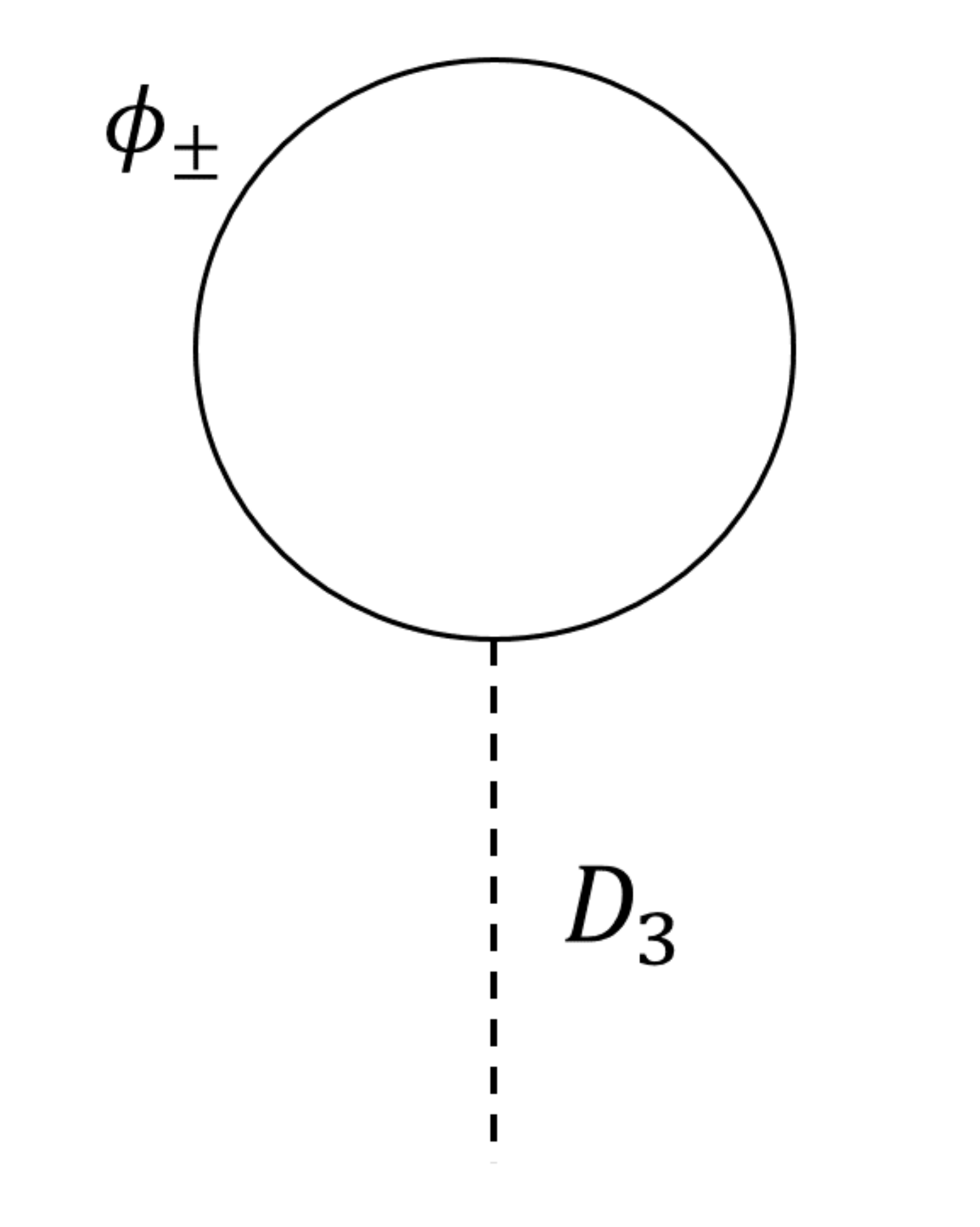}
\caption{The loop diagram that generates the FI-term.}
\label{fig:tadpole}
\end{figure}

We calculate the FI-term induced by 1-loop corrections of the scalar fields $\phi_\pm$.
As the first step, we use the mode expansions in the SUSY vacuum with $\xi=0$. 
In the SUSY vacuum with $\xi=0$, the EOMs (\ref{zero_mode_EOM}) and (\ref{massive_mode_EOM}) become
\begin{align}
	\partial \bar{\partial} \phi_\pm(z,\bar{z}) = \frac{R^2}{4} \lambda \phi_\pm(z,\bar{z}) ,
\end{align}
where we represent them in the complex coordinates $(z, \, \bar{z})$.
The general solutions of EOMs are given by
\begin{align}
	\phi_\pm (z,\bar{z}) = A \, e^{c z - c' \bar{z}},
\end{align}
where $A$ is a complex constant, and  $c,\,c'$ are also complex constants satisfying 
\begin{align}
	cc' = -\frac{R^2}{4} \lambda.
\end{align}
By imposing the boundary conditions $\phi_{\pm}(z+2\pi)= \phi_{\pm}(z)$
and $\phi_{\pm}(z+2\pi \tau)= \phi_{\pm}(z)$
the complex constants $c,\, c'$ are quantized:
\begin{align}
	&2\pi(c-c') = 2\pi i n~~(n\in \mathbb{Z}) ,\\
	&2\pi (c\tau - c'\bar{\tau}) =2\pi i \ell~~(\ell \in \mathbb{Z}) .
\end{align}
Thus the solutions that satisfy the boundary conditions are represented as follows:
\begin{align}
	&\phi_{\pm,n\ell}(z,\bar{z}) = A_{n\ell} e^{\frac{1}{2\text{Im}\, \tau} \big(  n(\tau\bar{z} -\bar{\tau }z)+\ell(z-\bar{z}) \big)} 
, \\
	&\lambda = -\frac{1}{R^2 (\text{Im}\, \tau)^2} \Big\{ (n\, \text{Re}\, \tau -\ell)^2 + (n\, \text{Im}\, \tau)^2 \Big\}.
\end{align}
In the coordinates $(x'_5,\, x'_6)$, these can be more simple form as
\begin{align}
	\phi_{\pm,n\ell}(x'_5, \,x'_6) = A_{n\ell} \,e^{i(\frac{n}{R}x'_5 + \frac{\ell}{R}x'_6)}.
\end{align}
Since $(n,\ell)$ and $(-n,-\ell)$ correspond to the same eigenvalue,
\begin{align}
\phi_{\pm,n\ell} (x'_5,\,x'_6)  &= 
A_{n\ell} e^{i(\frac{n}{R}x'_5 + \frac{\ell}{R}x'_6)}  + B_{n\ell} e^{-i(\frac{n}{R}x'_5 + \frac{\ell}{R}x'_6)},
\end{align}
where $n$ runs from 0 to $+\infty$ and $\ell$ runs from $-\infty$ to $+\infty$.
Under the action of $Z_2$, the wave functions behave as
\begin{align}
	&\phi_+(-x'_5,\,-x'_6) = \phi_+ (x'_5,\,x'_6), \\
	&\phi_-(-x'_5,\,-x'_6) = -\phi_- (x'_5,\,x'_6).
\end{align}
We obtain mode expansions of the bulk scalars:
\begin{align}
&\phi_{+,n\ell} (x'_5,\,x'_6) =A_\lambda \cos\left(\frac{n}{R}x'_5 + \frac{\ell}{R}x'_6\right),\\
&\phi_{-,n\ell} (x'_5,\,x'_6) =A_\lambda \sin\left(\frac{n}{R}x'_5 + \frac{\ell}{R}x'_6\right),
\end{align}
where the normalization factor $A_\lambda$ is $1/{\pi R \sqrt{\text{Im}\, \tau}}$ for $\lambda \neq 0$ up to phases, which are not 
relevant to the following discussions.
Zero modes are constant solutions.
They are given by
\begin{align}
	&\phi_{+,00} = A_0~~~~(A_0 = 1/2\pi R \sqrt{\text{Im}\, \tau}), \\
	&\phi_{-,00} = 0,
\end{align}
up to a phase, which is not 
relevant to the following discussions.
1-loop diagrams contributing to the FI-term are written as Figure \ref{fig:tadpole} in the case of $S^1/Z_2$.\footnote{The loop diagram around which the scalars $\phi_\pm$ run induces only the linear term of $D_3$.
The same contribution to the linear term of $F_{5'6'}$ arise from fermion's loop as same as the $\partial_y \Sigma$ in the $S^1/Z_2$ model unless the SUSY is broken.}
We can evaluate the divergent part of the FI-term that is induced by 1-loop diagrams of bulk scalars:
\begin{align}
	\xi_{\text{bulk}}(x'_5,\,x'_6) &= g\, \mbox{tr}(q) \Big( \frac{\Lambda^2}{16\pi^2}+ \frac{1}{4} \frac{\ln \Lambda^2}{16\pi^2}g^{i'j'} \partial_{i'} \partial_{j'}\Big) \sum_{n=0}^{\infty} 
\sum_{l =-\infty}^{\infty} \{  |\phi_{+,nl}|^2 -|\phi_{-,nl}|^2 \} \notag \\
&=g\, \mbox{tr}(q) \Big( \frac{\Lambda^2}{16\pi^2}+ \frac{1}{4} \frac{\ln \Lambda^2}{16\pi^2} g^{i'j'} \partial_{i'} \partial_{j'}\Big) \frac{1}{4|\text{Im}\, \tau|} \sum_{I=1,...,4} \delta(x'_5-x'^I_5)\delta(x'_6-x'^I_6),
\label{BulkFI_NoVEV_tilted}
\end{align}
where the second derivative $g^{i'j'} \partial_{i'} \partial_{j'}=\frac{4}{R^2} \partial \bar{\partial}$ is originated from the log divergent term $+\frac{1}{4}\lambda \ln \Lambda^2$ by use of the EOM.
In the second row, we use the Fourier expansion of the Dirac delta function:
\begin{align}
	\delta(y) = \frac{1}{\pi R} +\frac{2}{\pi R} \sum_{n>0}^{\infty} \cos\Big( \frac{2ny}{R} \Big) ~~(-\pi R<y< \pi R).
\end{align}
Note that the factor $1/|\text{Im}\, \tau|$ is multiplied,
which comes from $\sqrt{\text{det} g_{i'j'}}=|\text{Im}\, \tau|$ when we normalize the wave function.
Considering the contributions from the brane modes, we obtain the 1-loop induced FI-term:
\begin{align}
	\xi (x'_5,\,x'_6)&= \xi_{\text{bulk}}+\xi_{\text{brane}} \notag \\
	&= \frac{1}{|\text{Im}\, \tau|}\sum_{I=1,...,4} (\xi_I+\xi''g^{i'j'} \partial_{i'} \partial_{j'}) \delta(x'_5-x'^I_5)\delta(x'_6-x'^I_6),\label{1-loopFI:NoBG_1} \\
	&\hspace*{-45pt} 
\xi_I=g \frac{\Lambda^2}{16\pi^2} \Big( \frac{1}{4} \text{tr}(q) + \text{tr}(q_I)\Big) ,\,\, \xi'' = \frac{g}{4} \frac{\ln \Lambda^2}{16\pi^2}\,\frac{1}{4}\text{tr}(q). \label{1-loopFI:NoBG_2}
\end{align}
The FI-term is localized at the fixed points of the orbifold.
Thus we obtain a localized FI-term.

\subsection{Zero Mode when $\xi\neq0$}
\label{Sec:Background}

On the untilted torus, i.e., $\text{Re}\, \tau =0$, the zero mode of scalar field is localized at the fixed points by the localized FI-term \cite{Lee:2003mc}.
Here, we show that the FI-term localizes the zero mode of scalar field similarly  at the fixed points in the general $T^2/Z_2$ 
orbifold with arbitrary $\tau$.

From (\ref{SUSY_Cond'}) and (\ref{zero_mode_EOM'}), the SUSY conditions and the EOM of the zero mode for the bulk scalar are represented by
\begin{align}
	&\braket{F_{5'6'}} = |\text{Im}\, \tau|\, \xi (x'_5,\,x'_6), \label{F_SUSY_Condition} \\
	&(\tau \mathcal{D}_{5'}-\mathcal{D}_{6'}) \phi_{\pm,0}(x'_5,\,x'_6)=0.
\end{align}
We concentrate on the parity even mode.%
\footnote{Obviously the parity odd modes have no zero mode.}
We write explicitly them by the derivatives $\partial_{5'}, \partial_{6'}$ and gauge fields $A_{5'},\, A_{6'}$:
\begin{align}
	&\partial_{5'}\braket{A_{6'}} -\partial_{6'} \braket{A_{5'}} = |\text{Im}\, \tau| \xi(x_5',\, x_6'), \\
	&\Big\{(\tau \partial_{5'}-\partial_{6'})+igq(\tau\braket{ A_{5'}}-\braket{A_{6'}})\Big\} \phi_{+,0}(x_5',\, x_6') =0.
\end{align}
Here, we consider the following gauge fixing conditions:\footnote{Considering $\text{Re}\tau=0$ and the differences of scale between $x_6$ and $x'_6$, we see that this gauge (\ref{gauge}) intrinsically corresponds to the gauge in \cite{Lee:2003mc}.}
\begin{align}
\begin{cases} \label{gauge}
	A_{5'} =(\text{Im}\, \tau)^{-1} \big( \text{Re}\tau \,\partial_{5'} - \partial_{6'} \big) W,  \\
	A_{6'} =(\text{Im}\, \tau)^{-1} \big( |\tau|^2 \partial_{5'} -\text{Re}\tau \, \partial_{6'} \big) W.
\end{cases}
\end{align}
In this gauge,
the SUSY condition and EOM become
\begin{align}
	&\frac{1}{\text{Im}\, \tau} \big( |\tau|^2 \partial_{5'}^2 -2\text{Re}\, \tau \partial_{5'}\partial_{6'} +\partial_{6'}^2 \big) \braket{W} = |\text{Im}\, \tau| \xi(x_5',x_6'), \label{SUSY_Con1}\\
	&\Big\{(\tau \partial_{5'}-\partial_{6'})-gq (\tau \partial_{5'}-\partial_{6'}) \braket{W}\Big\} \phi_{+,0}(x_5',x_6') =0.\label{EOM1}
\end{align}
In the complex coordinates $Rz = x'_5+\tau x'_6$ and $R\bar{z} = x'_5 +\bar{\tau} x'_6$, the derivatives $\partial_z, \, \partial_{\bar{z}}$ are given by
\begin{align}
	\begin{pmatrix} \partial_z \\ \partial_{\bar{z}} \end{pmatrix} = -\frac{R}{\tau-\bar{\tau}}\begin{pmatrix} \bar{\tau} & -1 \\ -\tau & 1\end{pmatrix}  \begin{pmatrix} \partial_{5'} \\ \partial_{6'} \end{pmatrix}.
\end{align}
Eqs.~(\ref{SUSY_Con1}) and (\ref{EOM1}) are written as follows:
\begin{align}
	&  \partial \bar{\partial} \braket{W} =\frac{R^2}{4} \xi, \label{N_gauge1}\\
	&\big\{\bar{\partial}-gq (\bar{\partial} \braket{W})\big\} \phi_{+,0}(z,\bar{z}) =0, \label{N_gauge2}
\end{align}
where the 1-loop FI-terms (\ref{1-loopFI:NoBG_1}) and (\ref{1-loopFI:NoBG_2}) 
are represented in the complex coordinate as
\begin{align}
	&\xi(z,\bar{z}) = \frac{2}{R^2}\sum_{I=1,...,4} (\xi_I +\xi'' \frac{4}{R^2}\partial \bar{\partial} ) \delta^{(2)}(z-z_I), \label{FI-term_NoBackground_1}\\
	&\xi_I = g\frac{\Lambda}{16\pi^2}(\frac{1}{4} \text{tr}(q) +\text{tr} (q_I)), \,\, \xi'' = \frac{g}{4} \frac{\ln \Lambda^2}{16\pi^2}\frac{1}{4}\text{tr}(q),\label{FI-term_NoBackground_2}
\end{align}
where the factors come from the coordinate transformation.
From (\ref{N_gauge1}) and (\ref{FI-term_NoBackground_1}), we can split the SUSY solution into two parts:
\begin{align}
	&\braket{W}= \braket{W'}/2+\braket{W''}, \label{Split:W-background} \\
	\partial \bar{\partial}\braket{W'} = \sum_{I=1,...,4} \xi_I &\delta^{(2)}(z-z_I), \,\,\,\, \braket{W''} = \frac{2}{R^2}\sum_{I=1,...,4}\xi'' \delta^{(2)}(z-z_I). \label{Component:W-background}
\end{align}
The equation for $\braket{W'}$ is the Poisson equation 
with the source $\xi_I$ at the fixed points.
The solution is obtained as
\begin{align}
	\braket{W'} = \frac{1}{2\pi} \sum_I \xi_I \bigg[ \ln \Big| \vartheta_1\Big(\frac{z-z_I}{2\pi}\Big|\tau\Big) \Big|^2 -\frac{1}{2\pi \text{Im}\, \tau} \{\text{Im}\, (z-z_I) \}^2 \bigg].
\label{eq:W-backgroud}
\end{align}
Here $\vartheta_1(z|\tau)$ is the elliptic theta function, and our convention is given by
\begin{align}
	&\vartheta_{ab}(z,\tau) = \sum_{n=-\infty}^{\infty} e^{\pi i (n+a/2)^2\tau +2\pi i (n+a/2)(z+b/2)},\\
	\vartheta_1(z|\tau) \equiv -\vartheta_{11}(z,\tau),& \,\, \vartheta_2(z|\tau) \equiv \vartheta_{10}(z,\tau),\,\, \vartheta_3(z|\tau) \equiv \vartheta_{00}(z,\tau),\,\, \vartheta_4(z|\tau) \equiv \vartheta_{01}(z,\tau).
\end{align}
With this gauge background, the solution of the EOM (\ref{N_gauge2}) can be formally represented by
\begin{align}
	\phi_{+,0}(z,\bar{z}) = f(z) e^{gq \braket{W}}.
\end{align}
The holomorphic function $f(z)$ must be constant because it is a periodic holomorphic function.
The zero mode of $\phi_+$ is represented as follows:
\begin{align}
	\phi_{+,0}(z,\bar{z}) = f \prod_{I= 1...4} \Big| \vartheta_1\Big(\frac{z-z_I}{2\pi}\Big| \tau \Big) \Big|^{gq\xi_I/2\pi} \exp\Big\{ -\frac{gq \xi_I}{8\pi^2 \text{Im}\, \tau}\{\text{Im}\, (z-z_I)\}^2+\frac{2gq \xi''}{R^2}\delta^{(2)}(z-z_I) \Big\}. \label{even_zero_mode}
\end{align}
Since this wave function includes the Dirac delta function in the argument of exponential, 
it is not well defined.
The Dirac delta function implies that this wave function has serious divergences at the fixed points,
while the fixed points are the zero points for the theta function.
Integral of the wave function on any small region including a fixed point seems to be divergent.
Whereas wave functions must be canonically normalized.
This divergence must be canceled by the normalization factor $f$.
As a result, normalized wave function would be a localized mode at the fixed points such as the Dirac delta function.
Such a localized mode appears in an explicit regularization scheme for the case of $S^1/Z_2$ compactification \cite{GrootNibbelink:2002qp, Abe:2019nkv}.
It is also true for toroidal orbifolds.
We can show it by use of an explicit regularization of the delta function.%
\footnote{See Appendix \ref{app:local}.}

\subsection{1-loop FI-term when $\xi\neq0$}
\label{deformation of phi}

Calculation of the 1-loop FI-term is affected by the zero mode localization.
It implies the instability of the supersymmetric vacuum for the $S^1/Z_2$ model \cite{Abe:2019nkv}.
Such a vacuum instability may happen in the present $T^2/Z_2$ model.
Thus we should reevaluate the 1-loop FI-term again with the background given by \eqref{Split:W-background}, \eqref{Component:W-background} and \eqref{eq:W-backgroud}, and we should examine how stable configurations for the brane mode are.

In our evaluation, we make the following two assumptions:\\

\begin{tabular}{cl}
{\bf Assumption 1}: &The massive mode profiles of the bulk scalar are excluded at the \\
						&\ fixed points. \\
{\bf Assumption 2}: &Corrections to the FI-term can be evaluated by the square values of \\
						&\ wave functions near the fixed points only.
\end{tabular}
\newline\newline
The first assumption means that the induced FI-term can be evaluated by the zero mode of the bulk scalar field only.
It is true for the $S^1/Z_2$ model \cite{Abe:2019nkv}.%
\footnote{The massive mode of the bulk scalar field is evaluated in \cite{Lee:2003mc}. 
The evaluation was performed except the small regions that contain the fixed points, and the analysis near the fixed points are difficult.}
The second assumption means that the ratio of the 1-loop FI-term at each fixed point $z=z_I$ is equal to the ratio of $|\phi_{+,0}(z_I+\epsilon, \bar{z}_I +\bar \epsilon)|^2$ of \eqref{even_zero_mode}.%
\footnote{Since the zero mode wave function is localized at the fixed points, this description is not exactly true.
We provide a more rigorous treatment and justify the second assumption in Appendix \ref{app:local}.}

From (\ref{even_zero_mode}), the zero mode 
near the fixed point $z=z_I$ is written as below:
\begin{align}
	\phi_{+,0}(z_I+\epsilon, \bar{z}_I +\bar \epsilon) = 
	&f \exp \Big\{\frac{2gq\xi''}{R^2}\delta_\rho^{(2)}(\epsilon)\Big\} |\vartheta_1\left(\frac{\epsilon}{2\pi}\Big|\tau\right)|^{gq\xi_I/2\pi}\notag \\
	&\times \prod_{J \neq I} \Big| \vartheta_1\Big(\frac{z_I+\epsilon-z_J}{2\pi}\Big|\tau \Big) \Big|^{gq\xi_J/2\pi} \exp\Big\{ -\frac{gq \xi_J}{8\pi^2 \text{Im}\, \tau}\{\text{Im}\, (z_I +\epsilon -z_J)\}^2 \Big\},
\end{align}
where we introduce $\delta_\rho^{(2)}(z)$, which is a regularization of delta function; 
$\delta_\rho^{(2)}(z)$ is finite and $\delta_\rho^{(2)}(z) \rightarrow \delta^{(2)}(z)$ as $\rho \rightarrow +0$.\footnote{For a concrete example, see Appendix \ref{app:local}.}
We introduce $\xi_{\text{min}}$, which denotes the minimum of $\xi_I$.
$\vartheta_1(\epsilon/2\pi|\tau)$ is approximated by $\eta(\tau)^3 \epsilon$ near the origin \cite{Lee:2003mc},
where $\eta(\tau)$ is the Dedekind eta function.
We find $\vartheta_1(\epsilon/2\pi|\tau)\rightarrow 0$ in the limit of $\epsilon \rightarrow 0$.
We redefine the normalization factor by
\begin{align}
	f'\equiv f \exp \Big\{\frac{2gq\xi''}{R^2} \delta_\rho^{(2)}(\epsilon)\Big\} \left|\vartheta_1\left(\frac{\epsilon}{2\pi}\Big|\tau\right)\right|^{gq\xi_{\text{min}}/2\pi}.
\end{align}
$f'$ is a finite constant.
The zero mode near the fixed point is represented as
\begin{align}
	\phi_{+,0}(z_I+\epsilon,\bar{z}_I+\bar \epsilon) = &f' \left|\vartheta_1\left(\frac{\epsilon}{2\pi}\Big|\tau\right)\right|^{\frac{gq(\xi_I-\xi_{\text{min}})}{2\pi}}   
\prod_{J \neq I} 
\Big| \vartheta_1\Big(\frac{z_I + \epsilon -z_J}{2\pi} \Big|\tau \Big) \Big|^{gq\xi_J/2\pi} 
\nonumber
\\
&\times\exp\Big\{ -\frac{gq \xi_J}{8\pi^2 \text{Im}\, \tau}\{\text{Im}\, (z_I + \epsilon -z_J)\}^2 \Big\}.\notag
\end{align}
If $\xi_I$ is not equal to $\xi_{\text{min}}$, because of the suppression of $ |\vartheta_1(\epsilon/2\pi|\tau)|$, the wave function must vanish near the fixed point $z=z_I$.
Thus the part $ |\vartheta_1(\epsilon/2\pi)|^{gq(\xi_I-\xi_{\text{min}})/2\pi} $ determines the point where the zero mode is localized.
For instance, if $\xi_{I^*}$ is the only minimum
and $\xi_{J\neq I^*} > \xi_{I^*}$, 
the wave function is localized only at $z_I^*$.
Thus it is represented as
\begin{align}
\phi_{+,0} = \sqrt{\delta(z-z_{I^*})},
\end{align}
where the square root of the delta function denotes
that the wave function is localized at the fixed point $z_{I^*}$ and canonically normalized.
If several $\xi_I$ are the minimum simultaneously,
the zero mode is localized at the several fixed points $z_I$ where $\xi_I = \xi_{min}$.

The ratio of the zero mode of bulk scalar fields at the fixed points can be practically evaluated by
\begin{align}
	r_I \equiv \prod_{J
\neq I
} \Big| \vartheta_1\Big(\frac{z_I-z_J}{2\pi} \Big) \Big|^{gq\xi_J/2\pi} \exp\Big\{ -\frac{gq \xi_J}{8\pi^2 \text{Im}\, \tau}\{\text{Im}\, (z_I-z_J)\}^2 \Big\}.
\label{eq:r_I}
\end{align}
In the complex coordinates $(z,\bar{z})$, the fixed points are
\begin{align}
	z_I =\{0, \, \pi , \, \pi \tau , \, \pi (1+\tau)\}. \label{c.f.p.}
\end{align}
The explicit forms of $\{\text{Im}\, (z_I-z_J)\}^2$ are summarized in Table \ref{tab:z_I-z_J}.
\begin{table}
\centering
\begin{tabular}{|c||c|c|c|c|} \hline
$\{\text{Im}\, (z_I-z_J)\}^2$ & $J=1$ & $J=2$ & $J=3$ & $J=4$ \\
\hline \hline
$I=1$ & 0 & 0 & $\pi^2 (\text{Im}\, \tau)^2$ & $\pi^2 (\text{Im}\,\tau)^2$ \\
 \hline
$I=2$ & 0 & 0 & $\pi^2 (\text{Im}\,\tau)^2$ & $\pi^2 (\text{Im}\,\tau)^2$ \\ 
\hline
$I=3$ & $\pi^2 (\text{Im}\,\tau)^2$ & $\pi^2 (\text{Im}\,\tau)^2$ & 0 & 0 \\ 
\hline
$I=4$ & $\pi^2 (\text{Im}\,\tau)^2$ & $\pi^2 (\text{Im}\,\tau)^2$ & 0 & 0 \\ 
\hline
\end{tabular}
\caption{$\{ {\rm Im}\, (z_I-z_J)\}^2$.}
\label{tab:z_I-z_J}
\end{table}
We define $T_I$ as
\begin{align}
T_I  \equiv \prod_{J
\neq I} \Big| \vartheta_1\Big(\frac{z_I-z_J}{2\pi} \Big) \Big|^{gq\xi_J/2\pi}, 
\end{align}
which is the elliptic theta function part of $r_I$.
From (\ref{c.f.p.}), we find 
\begin{eqnarray}
T_I=
\left(
\begin{array}{ccccccccc}
\{ &\fs 1 & \fs \times &\fs  |\vartheta_1(-\frac{1}{2}|\tau)|^{\xi_2} &\fs  \times &\fs  |\vartheta_1(-\frac{\tau}{2}|\tau)|^{\xi_3} &\fs  \times&\fs |\vartheta_1(-\frac{1+\tau}{2}|\tau)|^{\xi_4} &\fs  \}^{gq/2\pi} \\
\{ &\fs |\vartheta_1(\frac{1}{2}|\tau)|^{\xi_1}&\fs  \times &\fs  1 &\fs  \times &\fs  |\vartheta_1(\frac{1-\tau}{2}|\tau)|^{\xi_3} &\fs \times& \fs |\vartheta_1(-\frac{\tau}{2}|\tau)|^{\xi_4} &\fs \}^{gq/2\pi} \\
\{ &\fs |\vartheta_1(\frac{\tau}{2}|\tau)|^{\xi_1}&\fs  \times &\fs  |\vartheta_1(-\frac{1-\tau}{2}|\tau)|^{\xi_2} &\fs  \times &\fs  1 &\fs \times&\fs |\vartheta_1(-\frac{1}{2}|\tau)|^{\xi_4} &\fs \}^{gq/2\pi} \\
\{ &\fs |\vartheta_1(\frac{1+\tau}{2}|\tau)|^{\xi_1}&\fs  \times &\fs  |\vartheta_1(\frac{\tau}{2}|\tau)|^{\xi_2} & \fs \times &\fs  |\vartheta_1(\frac{1}{2}|\tau)|^{\xi_3} &\fs \times&\fs 1 &\fs \}^{gq/2\pi}
\end{array}
\right), \notag
\end{eqnarray}
where the first, second, third and fourth rows correspond to $T_1$, $T_2$, $T_3$ and $T_4$ respectively.
The elliptic theta function $\vartheta_1$ satisfies the following relations:
\begin{align}
	&\vartheta_1(v+1|\tau)= -\vartheta_1(v|\tau), \\
	&\vartheta_1(v+\tau|\tau)= -e^{-i\pi (2v+\tau)} \vartheta_1(v|\tau),
\end{align}
and the elliptic theta functions $\vartheta_i(i=2,3,4)$ are related to $\vartheta_1$ as
\begin{align}
	&\vartheta_1\Big(\frac{1}{2}\Big|\tau \Big)=\vartheta_2(0|\tau), \\
	&\vartheta_1\Big(\frac{\tau}{2}\Big|\tau\Big)= i e^{-i\pi\tau/4} \vartheta_4(0|\tau), \\
	&\vartheta_1\Big(\frac{1+\tau}{2}\Big|\tau\Big)= \vartheta_2\Big(\frac{\tau}{2}\Big|\tau \Big)  = e^{-i\pi \tau/4} \vartheta_3(0|\tau).
\end{align}
Therefore, by using $\vartheta_i(0|\tau)\,(i=2,3,4)$, $T_I$ is simply rewritten as
\begin{eqnarray}
T_I =
\left(
\begin{array}{ccccccccc}
\{ & \fs |\vartheta_2(0|\tau)|^{\xi_2} &  \fs \times &\fs |\vartheta_3(0|\tau)|^{\xi_4} &  \fs \times &  \fs |\vartheta_4(0|\tau)|^{\xi_3} &  \fs \times&  \fs e^{\frac{\pi \text{Im}\,\tau}{4}(\xi_3+\xi_4)} &\fs \}^{gq/2\pi} \\
\{ &\fs |\vartheta_2(0|\tau)|^{\xi_1} & \fs \times &\fs |\vartheta_3(0|\tau)|^{\xi_3} &\fs \times & \fs |\vartheta_4(0|\tau)|^{\xi_4} &  \fs \times& \fs e^{\frac{\pi \text{Im}\,\tau}{4}(\xi_3+\xi_4)}  &\fs  \}^{gq/2\pi} \\
\{ & \fs |\vartheta_2(0|\tau)|^{\xi_4} & \fs \times & \fs |\vartheta_3(0|\tau)|^{\xi_2} & \fs \times &\fs |\vartheta_4(0|\tau)|^{\xi_1} & \fs \times& \fs e^{\frac{\pi \text{Im}\,\tau}{4}(\xi_1+\xi_2)} &\fs \}^{gq/2\pi} \\
\{ &\fs |\vartheta_2(0|\tau)|^{\xi_3} & \fs \times &\fs |\vartheta_3(0|\tau)|^{\xi_1} &  \fs \times & \fs |\vartheta_4(0|\tau)|^{\xi_2} &\fs \times& \fs e^{\frac{\pi \text{Im}\,\tau}{4}(\xi_1+\xi_2)} & \fs \}^{gq/2\pi}
\end{array}
\right).
\end{eqnarray}
The ratio of the absolute value of the wave functions at the fixed points is evaluated as
\begin{eqnarray}
r_I   = 
\left(
\begin{array}{ccccccc}
\{ &\fs |\vartheta_2(0|\tau)|^{\xi_2} &\fs \times &\fs |\vartheta_3(0|\tau)|^{\xi_4} &\fs  \times &\fs  |\vartheta_4(0|\tau)|^{\xi_3} &\fs \}^{gq/2\pi} \\
\{ &\fs |\vartheta_2(0|\tau)|^{\xi_1} &\fs \times &\fs |\vartheta_3(0|\tau)|^{\xi_3} &\fs  \times &\fs  |\vartheta_4(0|\tau)|^{\xi_4} &\fs \}^{gq/2\pi} \\
\{ &\fs |\vartheta_2(0|\tau)|^{\xi_4} &\fs \times &\fs |\vartheta_3(0|\tau)|^{\xi_2} &\fs  \times &\fs  |\vartheta_4(0|\tau)|^{\xi_1} &\fs \}^{gq/2\pi} \\
\{ &\fs |\vartheta_2(0|\tau)|^{\xi_3} &\fs \times &\fs |\vartheta_3(0|\tau)|^{\xi_1} &\fs  \times &\fs  |\vartheta_4(0|\tau)|^{\xi_2} &\fs \}^{gq/2\pi}
\end{array}
\right).
\label{zero mode density}
\end{eqnarray}
Since the zero mode is localized at the fixed points, the normalized wave function of the zero mode is given by
\begin{align}
|\phi_{+.0}| = \sqrt{\frac{\sum_{\xi_I =\xi_{min}}  r_I^2 \delta(z-z_I)}{\sum_{\xi_I =\xi_{min}} r_I^2}}.
\end{align}

$r_I$ are transformed each other by the modular symmetry.
The modular symmetry is generated by two elements, $S$ and $T$, 
and these generators transform the modulus $\tau$ as 
\begin{equation}
S:~\tau \rightarrow -\frac{1}{\tau}, \qquad T:~\tau \rightarrow \tau + 1.
\end{equation}
The elliptic theta functions are transformed each other by $S$ and $T$, 
and transformation behavior is shown in Appendix \ref{app:C}.
The $S$ transforms zero mode values at $z_1$ and $z_4$, and $z_2$ and $z_3$, i.e.,
$\phi_{+,0} (z_1,\bar{z}_1)  \longleftrightarrow \phi_{+,0} (z_4,\bar{z}_4) $ and 
$\phi_{+,0} (z_2,\bar{z}_2)  \longleftrightarrow \phi_{+,0} (z_3,\bar{z}_3) $.
On the other hand, the $T$ transforms zero mode values at  $z_1$ and $z_2$, and $z_3$ and $z_4$, i.e.,
$\phi_{+,0} (z_1,\bar{z}_1)  \longleftrightarrow \phi_{+,0} (z_2,\bar{z}_2) $ and 
$\phi_{+,0} (z_3,\bar{z}_3)  \longleftrightarrow \phi_{+,0} (z_4,\bar{z}_4) $.
When $\xi_1=\xi_2=\xi_3=\xi_4$, the above zero mode profile is invariant under the 
modular symmetry.

\section{Stability of SUSY vacua on untilted torus}

In the previous section, we have finished the preparations to calculate the localized FI-term in the new SUSY background, where the VEV of $F_{5'6'}$ has nonzero value.
In stable configuration,
the bulk mode contribution cancels the brane mode contributions. 
Thus we examine configurations where the cancellation occurs.
Under the second assumption, the 1-loop FI-term that is induced by the bulk mode can be evaluated by $r_I^2$.
In the configurations where the cancellation cannot occur, the 1-loop FI-term changes the supersymmetric vacuum further, which leads to the instability of the SUSY vacuum.

In this section, we investigate the stability of the SUSY vacuum in the untilted torus, i.e., $\text{Re}\, \tau =0$.
In the untilted torus, except for the differences from the scale of $x_6$ and $x'_6$, the zero mode profile $\phi_{+,0}$ and gauge field $W$ coincide with the results in \cite{Lee:2003mc}.

\subsection{Completely symmetric configuration}\label{Completely_Symmetric}
First we consider the completely symmetric configuration of the brane charges, i.e., $q_1=q_2 = q_3 =q_4$. We assume the sum of $U(1)$ charges is set to zero, which means that the bulk charge is four times as big as that of the localized charge: $q=-4q_1$.
Furthermore, we assume the tree level Lagrangian has no FI-term and $\braket{F_{5'6'}}=0$. 
From (\ref{1-loopFI:NoBG_1}) and (\ref{1-loopFI:NoBG_2}), we obtain the 1-loop induced FI-term:
\begin{align}
\xi 
&= \xi_{\text{bulk}}+\xi_{\text{brane}} \notag \\
&= \frac{2}{R^2}\sum_{I=1,...,4} (\xi_I+\xi''\frac{4}{R^2}\partial \bar{\partial}) \delta^{(2)}(z-z_I), \\
&\xi_1=\xi_2=\xi_3=\xi_4=0 ,\,\, \xi'' = \frac{gq}{16} \frac{\ln \Lambda^2}{16\pi^2}.
\end{align}
Solving the D-flat condition \eqref{F_SUSY_Condition} in the gauge \eqref{gauge}, we obtain the corrected SUSY background solution:
\begin{align}\label{CS_bulk}
	&\braket{W}= \frac{2}{R^2}\sum_{I=1,...,4}\xi'' \delta^{(2)}(z-z_I).
\end{align}
In this new SUSY background, we recompute the zero mode of $\phi_+$.
The zero mode can be evaluated from (\ref{zero mode density}):
\begin{align}
	\phi_{+,0}(z,\bar{z}) = \frac{\sqrt{2}}{R}\sqrt{\frac{1}{4}\sum_{I=1,...,4}\delta^{(2)}(z-z_I) },
\label{eq:zero_part_sym}
\end{align}
where 
the square root of the delta function denotes
that the wave function is localized at the fixed points and canonically normalized as mentioned before.

Substituting \eqref{eq:zero_part_sym} into the KK expansion of the bulk fields in (\ref{BulkFI_NoVEV_tilted}), we obtain the 1-loop FI-term again.
From the assumption 1 in section \ref{deformation of phi}, the massive modes do not contribute to the 1-loop FI-term.
We can evaluate the contribution of the bulk fields:
\begin{align}
	\xi_{\text{bulk}}=gq \frac{\Lambda^2}{16\pi^2}\frac{1}{4}  \frac{2}{R^2}\sum_{I=1,...,4} \delta^{(2)}(z-z_I).
\end{align}
The contribution of the brane fields is unchanged.
It is written as
\begin{align}
	\xi_{\text{brane}} = g\frac{\Lambda^2}{16\pi^2} \frac{2}{R^2}\sum_{I=1,...,4} q_I \delta^{(2)}(z-z_I).
\end{align}
As a result, we obtain the quantum correction to the FI-term in the new SUSY background,
\begin{align}
	\xi(z, \bar{z}) = \xi_{\text{bulk}} + \xi_{\text{brane}} =0.
\end{align}
The quantum correction vanishes.
The bulk zero mode shields the brane charges completely.
Thus the SUSY vacuum does not shift further, i.e., it is a stable vacuum.

\subsection{Partially symmetric configuration}\label{Partially_Symmetric}

Next, we consider a partially symmetric configuration where the $U(1)$ charges of the brane fields are given by $q_1=0$ and $q_2 = q_3 =q_4$.
We assume the sum of $U(1)$ charges is set to zero, which means that the bulk charge is three times as big as that of the localized charge: $q=-3q_2$.
Furthermore, we assume the tree level Lagrangian has vanishing FI-term and $\braket{F_{5'6'}}=0$. 
The 1-loop induced FI-term is calculated as
\begin{align}
	\xi &= \xi_{\text{bulk}}+\xi_{\text{brane}} \notag \\
	&= \frac{2}{R^2}\sum_{I=1,...,4} (\xi_I+\xi''\frac{4}{R^2}\partial \bar{\partial}) \delta^{(2)}(z-z_I), \\
	 \xi_1 &=\kappa, \, \xi_2=\xi_3=\xi_4=-\kappa/3 \hspace*{10pt}\left(\kappa \equiv \frac{1}{4}gq \frac{\Lambda^2}{16\pi^2}\right), \\
\xi'' &= \frac{gq}{16} \frac{\ln \Lambda^2}{16\pi^2}.
\end{align}
Solving the D-flat condition \eqref{F_SUSY_Condition} in the gauge \eqref{gauge}, 
we obtain the SUSY background solution corrected by 1-loop effects as
\begin{align}\label{CS_bulk}
	&\braket{W}= \frac{1}{4\pi} \sum_{I=1,...,4} \xi_I \bigg[ \ln \Big| \vartheta_1\Big(\frac{z-z_I}{2\pi}\Big|\tau\Big) \Big|^2 -\frac{1}{2\pi \text{Im}\,\tau} \{\text{Im}\,(z-z_I) \}^2 \bigg]+ \frac{2}{R^2}\sum_{I:f.p.}\xi'' \delta^{(2)}(z-z_I).
\end{align}
The ratio of the zero mode at the fixed points in this new background can be evaluated from (\ref{zero mode density}):
\begin{eqnarray}
r_I = 
\left(
\begin{array}{ccccccc}
&&&0&&& \\
\{ &|\vartheta_2(0|\tau)|^{\kappa} & \times & |\vartheta_3(0|\tau)|^{-\kappa/3} & \times &|\vartheta_4(0|\tau)|^{-\kappa/3}  &\}^{gq/2\pi} \\
\{ &|\vartheta_2(0|\tau)|^{-\kappa/3} & \times & |\vartheta_3(0|\tau)|^{-\kappa/3} & \times & |\vartheta_4(0|\tau)|^{\kappa} &\}^{gq/2\pi} \\
\{ &|\vartheta_2(0|\tau)|^{-\kappa/3} & \times &|\vartheta_3(0|\tau)|^{\kappa} & \times & |\vartheta_4(0|\tau)|^{-\kappa/3}  &\}^{gq/2\pi}
\end{array}
\right). \notag
\end{eqnarray}
Note that the wave function of the zero mode vanishes at $z_1$ since $\xi_1$ is bigger than $\xi_{min} = -\kappa/3$.
The zero mode is given by
\begin{align}
	\phi_{+,0}(z,\bar{z}) =\frac{\sqrt{2}}{R} \sqrt{\frac{ |\vartheta_2(0|\tau)|^{\frac{4gq\kappa}{3\pi}} \delta^{(2)}(z-z_2)+|\vartheta_4(0|\tau)|^{\frac{4gq\kappa}{3\pi}} \delta^{(2)}(z-z_3)+|\vartheta_3(0|\tau)|^{\frac{4gq\kappa}{3\pi}} \delta^{(2)}(z-z_4)}{|\vartheta_2(0|\tau)|^{4gq\kappa/3\pi}+|\vartheta_3(0|\tau)|^{4gq\kappa/3\pi}+|\vartheta_4(0|\tau)|^{4gq\kappa/3\pi}}}.
\label{eq:phi_not1}
\end{align}

Substituting \eqref{eq:phi_not1} into the KK expansion of the bulk fields in (\ref{BulkFI_NoVEV_tilted}), we obtain the 1-loop FI-term again.
The contribution of the bulk field is given by
\begin{align}
	\xi_{\text{bulk}} &=gq \frac{\Lambda^2}{16\pi^2}\frac{2}{R^2} \notag \\ &\hspace*{10pt}\times \frac{ |\vartheta_2(0|\tau)|^{\frac{4gq\kappa}{3\pi}} \delta^{(2)}(z-z_2)+|\vartheta_4(0|\tau)|^{\frac{4gq\kappa}{3\pi}} \delta^{(2)}(z-z_3)+|\vartheta_3(0|\tau)|^{\frac{4gq\kappa}{3\pi}} \delta^{(2)}(z-z_4)}{|\vartheta_2(0|\tau)|^{4gq\kappa/3\pi}+|\vartheta_3(0|\tau)|^{4gq\kappa/3\pi}+|\vartheta_4(0|\tau)|^{4gq\kappa/3\pi}}.
\end{align}
The contribution of the brane fields is unchanged, and is written as
\begin{align}
	\xi_{\text{brane}} = g\frac{\Lambda^2}{16\pi^2}\frac{2}{R^2}\sum_{I=1,...,4} q_I \delta^{(2)}(z-z_I).
\end{align}
As a result, we obtain the quantum correction to the FI-term in the new SUSY background,
\begin{align}
	\xi(z,\bar{z}) = \xi_{\text{bulk}} + \xi_{\text{brane}} \neq 0.
\end{align}
The quantum correction does not vanish.
Therefore, the SUSY vacuum shifts further by the 1-loop FI-term, i.e., it is an unstable vacuum.
Unless we introduce a fine-tuned FI-term at tree level, the vacuum is unstable in the partially symmetric configuration.

\subsection{Stable and unstable configurations}
\label{Result_T}
We have examined the stability of the SUSY vacuum in the two configurations: completely symmetric one and partially symmetric one.
The former has the supersymmetric stable vacuum, 
but the latter does not.

\begin{table}[bt]\center
  \begin{tabular}{|ll|c|} \hline
   the charges of brane modes & & stability of the vacuum  \\ \hline \hline
   $ q_1=q_2=q_3=q_4$ & & stable  \\ \hline
   $q_1=q_2=q_3 \neq q_4$&  & unstable \\ \hline
   $q_1=q_2\neq q_3=q_4 $&  & stable \\ \hline
   $\{q_1=q_2\neq q_3,q_4 \}$ and $\{q_3\neq q_4\}$& & unstable\\ \hline
   $q_I \neq q_J\, (I\neq J)$&  & unstable \\ \hline
  \end{tabular}
\caption{Stable and unstable configurations of brane modes.}
\label{tab:(un)stable_conf}
\end{table}

We summarize stability of various configurations in Table \ref{tab:(un)stable_conf}.
In all of these examples, we assume that the bulk mode has a charge $q$ which cancels the charges of the brane modes, i.e., $q+\sum_I q_I =0$.
The first and second rows  correspond to the results in the section \ref{Completely_Symmetric} and \ref{Partially_Symmetric}, respectively.
In the table, 
``stable'' means that the FI-term is not induced in the new SUSY vacuum.
On the other hand ``unstable'' means that the FI-term is induced in the new SUSY vacuum.
It is always possible to introduce a localized FI-term at tree level
which makes the zero mode wave function of the bulk field shield the brane charges completely.
If such a fine-tuned FI-term is available,  unstable configurations can be stabilized.
To add the tree level FI-term, we should pay attention for flux quantization.
The localized FI-term corresponds to localized magnetic flux\cite{Scrucca:2004jn, vonGersdorff:2006nt, Buchmuller:2018lkz}.
The Wilson loop around the fixed points in the SUSY background of  \eqref{F_SUSY_Condition} is non-trivial,
\begin{align}
W_I = \exp\left(-iq \oint_{\mathcal{C}_I} A \right) = \exp\left( -iq \int_{\mathcal{D}_I} \xi \right),
\end{align}
where $\mathcal{C}_I$ is a circle around $z_I$ and $\mathcal{D}_I$ is the disc including $z_I$, and 
we use EOM of the gauge field \eqref{SUSY_Cond}.
Thus $\xi$ can be interpreted as a localized flux.
Since $W_I$ must be $\pm 1$ \cite{Buchmuller:2018lkz},
tree level FI-term is not a free parameter.
It is not clear whether we can always put appropriate $\xi_I$ which make the localization of the zero mode shield the brane charges completely,
satisfying the quantization condition.
It might be interesting to investigate it.

Vacuum (in)stability will be also related
to the anomaly on the compact space.
We observe that the stable configurations are anomaly free since the charge of the bulk zero modes is canceled by that of the brane modes everywhere.
On the other hands, anomaly is not canceled in the unstable configurations locally.
This may imply inconsistency of the model.
The local anomaly requires additional fields, e.g., antisymmetric fields,
which cancel the anomaly via Green-Schwarz mechanism, or other local operators.
These additional terms may change the localized FI-term
and vacuum structure.
For instance, the loop diagrams including antisymmetric fields would contribute to the localized FI-term, and shift it.
It may be interesting to investigate stability of the bulk mode including such additional effects.
We would study it elsewhere.

\section{Stability of SUSY vacua on tilted torus}

We examine the stability of the SUSY vacuum in the tilted torus $T^2/Z_2$, i.e., $\text{Re}\, \tau \neq 0$.
Basically, the results 
are the same as those of the untilted torus.
The difference comes only from the profiles of the zero modes, which generally depend on the background geometry.
Taking into account general $\tau$,
we find a part of unstable vacuum can be stabilized.
Especially, the partially symmetric configuration leads to different results.

\subsection{Stable configuration and Moduli stabilization}
\label{Stable configuration and Moduli stabilization}

\begin{figure}[hbtp]
\begin{center}
\includegraphics[scale=0.455]{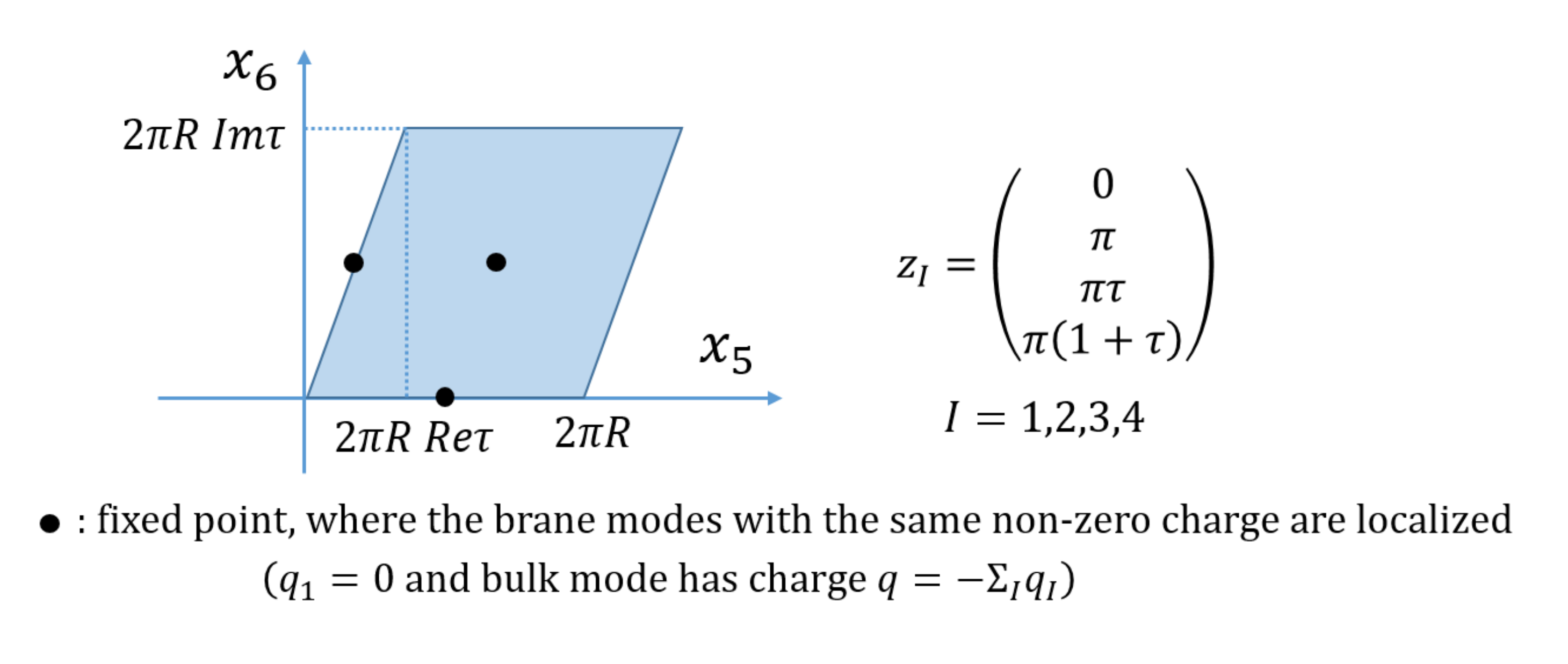}
\end{center}
\caption{The configuration of brane modes}
\label{config._brane_mode}
\end{figure}

We are interested in the partially symmetric configuration, i.e., the charges of three brane modes are the same,
and the charge of the other one is zero. 
Similar to section \ref{Partially_Symmetric}, we concentrate on the configuration that the charges of the brane modes in the fixed points $z=z_2,z_3,z_4$ are the same for concreteness.
The charge of the bulk mode is three times as big as that of the localized charge, which is required for $\sum_I \xi_I=0$. (See Figure \ref{config._brane_mode}.)

In the SUSY vacuum with $\braket{F_{5'6'}} =0$, the 1-loop induced FI-term is written by
\begin{align}
	\xi_1= \kappa,\, \, \xi_2 = \xi_3 =\xi_4=-\kappa/3 ,\,\, \xi'' \neq 0, \label{OriginalFI}
\end{align}
where $\kappa \equiv \frac{1}{4}gq \frac{\Lambda^2}{16\pi^2}$.

The FI-term corrects the SUSY vacuum as $\braket{F_{5'6'}} =|\text{Im}\,\tau|\, \xi (x'_5,\,x'_6)$. Again we evaluate the 1-loop FI-term in the new SUSY vacuum.
Since (\ref{OriginalFI}) satisfies $\xi_{\text{min}}=\xi_2=\xi_3=\xi_4$,
the $\phi_{+,0}(z_I,\bar{z}_I)$ is already given by \eqref{eq:phi_not1}. 
The ratio of zero mode profiles at fixed points is given by
\begin{align}
	|\phi_{+,0}(z_1)|^2:|\phi_{+,0}(z_2)|^2:|\phi_{+,0}(z_3)|^2:|\phi_{+,0}(z_4)|^2 = 0:|\vartheta_2(0|\tau)|^{\frac{4gq\kappa}{3\pi}}:|\vartheta_4(0|\tau)|^{\frac{4gq\kappa}{3\pi}}:|\vartheta_3(0|\tau)|^{\frac{4gq\kappa}{3\pi}}. \notag
\end{align}
The 1-loop FI-term induced by the bulk field in the new vacuum is induced as this ratio at the fixed points.
In order not to generate the 1-loop FI-term in the new vacuum,
the bulk contribution must cancel that from the brane modes.
Since the charges of the brane modes are the same at the three fixed points of $z_2,z_3,z_4$,
we obtain the following stability condition:\footnote{For other combinations of three fixed points where the three brane modes are located,
the equivalent conditions appear.}
\begin{align}
	|\vartheta_2(0|\tau)|=|\vartheta_3(0|\tau)|=|\vartheta_4(0|\tau)|. \label{Condition}
\end{align}
These conditions cannot be satisfied if $\text{Re}\, \tau=0$.
This is the reason why we insisted that this configuration is unstable in the untilted torus in section \ref{Result_T}. 
Whereas, in the tilted torus, the condition (\ref{Condition}) can be satisfied

By use of modular transformation behavior of the elliptic theta functions 
as shown in Appendix \ref{app:C}, 
we find that the complex structure, e.g., $\tau= e^{i\pi/3}$, satisfies the above condition (\ref{Condition}).
The point $\tau= e^{i\pi/3}$ is on the boundary of the fundamental domain of the modular group.
Thus, in the torus which has the complex structure $\tau = e^{i\pi/3}$, 
the 1-loop induced FI-term in the new SUSY vacuum vanishes.
Accordingly the configuration of three brane modes has a stable vacuum.

The 1-loop FI-term
generates a D-term potential:
\begin{align}
	V_D \propto \int dx'_5 dx'_6\sqrt{\text{det}g_{i'j'}}  (\xi+\cdots)^2.
\end{align}
$\xi$ contains the divergent term of cutoff $\Lambda$ if $\xi$ is not zero.
The D-term potential would be dominant.
Thus, we consider that 
$\tau$ would be stabilized in the value that cancels the 1-loop FI-term in the new SUSY vacuum.

\subsection{Stabilized complex structure}

In the configuration of three brane modes, we insist that the complex structure is stabilized dynamically at $\tau = e^{i\pi/3}$ by the potential $V_D$.

\begin{figure}[hbtp]
\begin{center}
\includegraphics[scale=0.4]{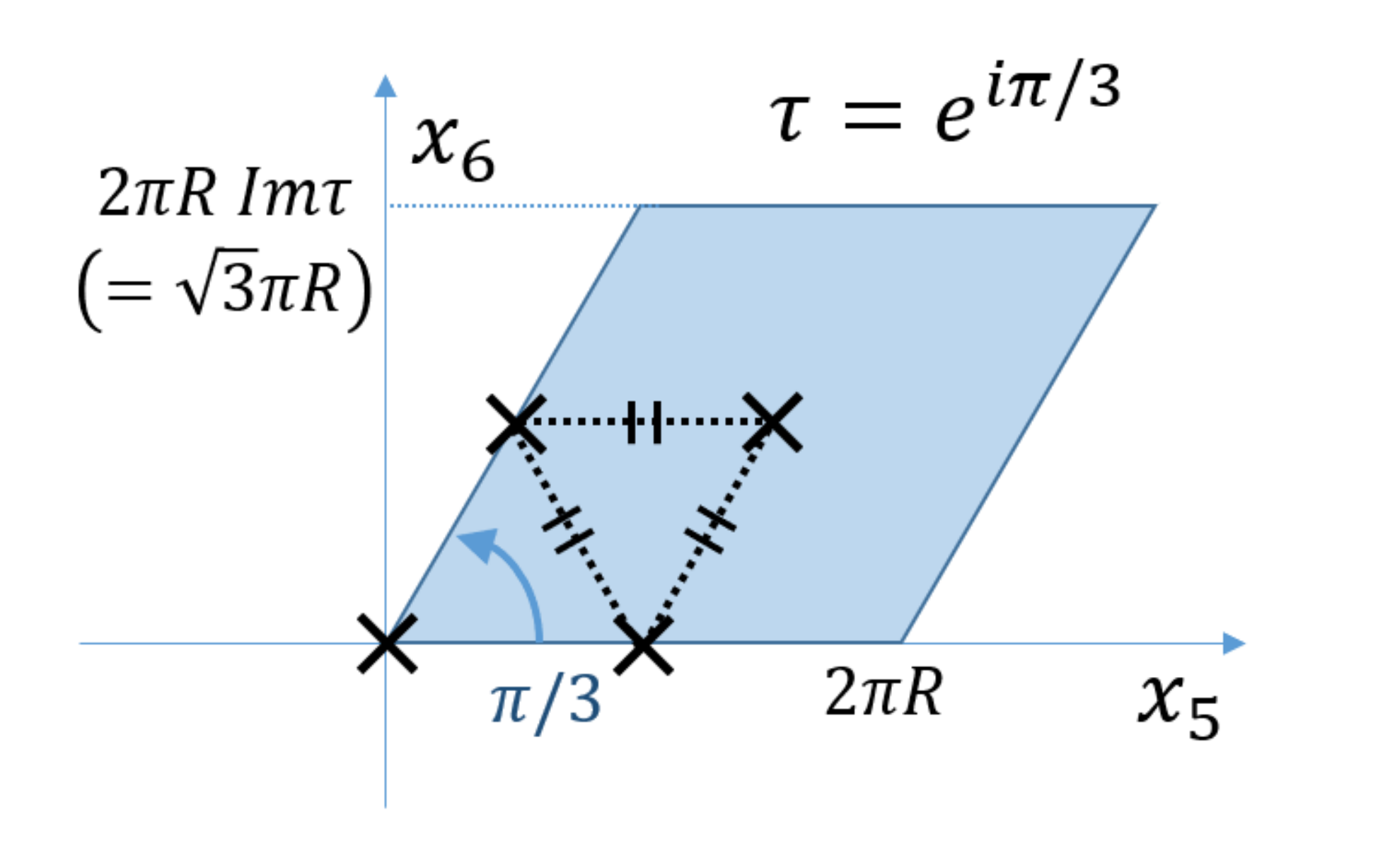}
\end{center}
\caption{Torus of $\tau = e^{i\pi/3}$}
\label{stable torus}
\end{figure}

We show the stable configuration in Figure \ref{stable torus}.
In this configuration, there are the brane modes in the fixed points except the origin,
and the bulk mode is localized at the fixed points except the origin, too.
Figure \ref{stable torus} shows  when the vacuum is stable,
the positional relations of fixed points where the branes are located
are equidistant each other.
We expect that the complex structure is stabilized in such a way 
that the fixed points where the branes are located
 have symmetric positional relations.
Otherwise there are no stable SUSY vacuum, and SUSY or gauge symmetry would be broken.

Four-dimensional CP can be embedded into proper Lorentz transformation in higher dimensional theory, 
where extra dimensions are also reflected \cite{
  Green:1987mn,Strominger:1985it,Dine:1992ya,Choi:1992xp,Lim:1990bp,Kobayashi:1994ks}.
For example, in six dimensional theory, four-dimensional CP is combined with the  reflection,
\begin{equation}
z \rightarrow -\bar z, 
\end{equation}
so as to be embedded into six-dimensional proper Lorentz transformation.
Under the above reflection, the modulus transforms 
\begin{equation}\label{eq:tau-CP}
\tau \rightarrow -\bar \tau .
\end{equation}
Thus, when $\text{Re}\, \tau =0$, CP is conserved.
For other values of $\text{Re}\, \tau $, CP can be broken.
Hence, the value $\tau = e^{i\pi/3}$ has implication in CP violation physics.\footnote{
If theory has modular symmetry, the transformation (\ref{eq:tau-CP}) is meaningful up 
to the modular symmetry.(See e.g. \cite{Baur:2019kwi,Novichkov:2019sqv,Kobayashi:2019uyt}.)
That implies that CP is conserved at the values of $\tau$ at the boundary of the fundamental domain including 
$\tau = e^{i\pi/3}$.}

\section{Conclusion}

We have investigated the quantum corrections to the localized FI-terms in six-dimensional SUSY Abelian gauge theory compactified on the $T^2/Z_2$ orbifold.

In the $S^1/Z_2$ orbifold, the localization of bulk zero mode causes the instability of the vacuum.
Similarly,
the bulk zero mode is localized
 in the untilted $T^2/Z_2$ model,  too \cite{Lee:2003mc}. 
We find
that the new supersymmetric vacuum which is changed by 1-loop FI-term can be unstable in untilted compactification.
The instability is related to the configuration of brane modes and their $U(1)$ charges.
We have shown that the 1-loop correction vanishes for the completely symmetric configurations, but it is not true for the asymmetric configurations.
It is because the zero mode profile and brane charges cancel each other for the former case, but it does not happen for the latter case.
Therefore, in the asymmetric configurations the vacuum receives further corrections and is unstable.
If we put a fine-tuned FI-term in the tree level Lagrangian, we can realize a stable vacuum even for asymmetric configuration.
In such a stable vacuum, zero mode profile shields the brane charges completely,
and their corrections are canceled each other.
This result is the same as the one derived on the $S^1/Z_2$ orbifold \cite{Abe:2019nkv}.

As opposed to the $S^1/Z_2$ orbifold, the complex structure exists in the $T^2/Z_2$ orbifolds.
The 1-loop FI-term depends on the complex structure, i.e., the complex structure associates with the instability of the vacuum.
Especially, we can stabilize the complex structure $\tau$ by using the cancellation of 1-loop FI-term that is induced in a new supersymmetric vacuum. 
We have considered the configuration with three brane modes that are located at
each of three fixed points and have the same charge.
We have found that the complex structure $\tau$ is stabilized at the value of $e^{i\pi/3}$, 
which makes the three fixed points equidistant each other.
We expect that the stabilization mechanism which is caused by the cancellation of 1-loop FI-term occurs in more general orbifolds, 
and the stabilized complex structures make the positions of fixed points  symmetric.
It  contrasts with the traditional moduli stabilization mechanism by three form flux \cite{ Gukov:1999ya, DeWolfe:2002nn, Taylor:1999ii}.\footnote{Toroidal orbifolds have K\"ahler moduli in general.
The effective potential of our model does not include the K\"ahler moduli, and its stabilization by the bulk instability is not realized.
We need another moduli stabilization mechanism such as non-perturbative effects for the K\"ahler moduli \cite{Kachru:2003aw}.
}
We have focused on 1-loop corrections and mainly investigated stable configurations in the present paper.
For unstable vacuum, SUSY or gauge symmetry would be broken, and higher loop correction might play important role.
It is interesting to consider these effects.
We will study it elsewhere.

Magnetic flux also affects the
profiles of the wave function of the bulk fields, and increase the number of the chiral zero modes \cite{Abe:2008fi, Abe:2013bca, Abe:2014noa, Kobayashi:2017dyu}.
It is interesting to extend our analysis to the $T^2/Z_2$ orbifolds with magnetic fluxes.
Its flavor structure would be different from that of magnetized orbifold models without FI-terms 
\cite{Abe:2012fj,Abe:2014vza,Kobayashi:2016qag}.
In magnetized orbifold models, zero modes transform each other under the modular symmetry 
\cite{Kobayashi:2018rad,Kobayashi:2018bff,Kariyazono:2019ehj}.
In addition, our FI-term has already non-trivial behavior under the modular symmetry.
Thus, it is interesting to study localized FI-terms from the viewpoint of modular flavor models \cite{Feruglio:2017spp} and 
their modulus stabilization \cite{Kobayashi:2019uyt,Kobayashi:2019xvz}.

\section*{Acknowledgments}
H. A. was supported in part by Waseda University Grant for Special Research Projects (Project number: 2019Q-027) and also supported by Institute for Advanced Theoretical and Experimental Physics, Waseda University.
T. K. was supported in part by MEXT KAKENHI Grant Number JP19H04605.

\appendix
\section{Localization of the zero mode}\label{app:local}

\begin{figure}[t]
\begin{center}
\includegraphics[scale=0.40]{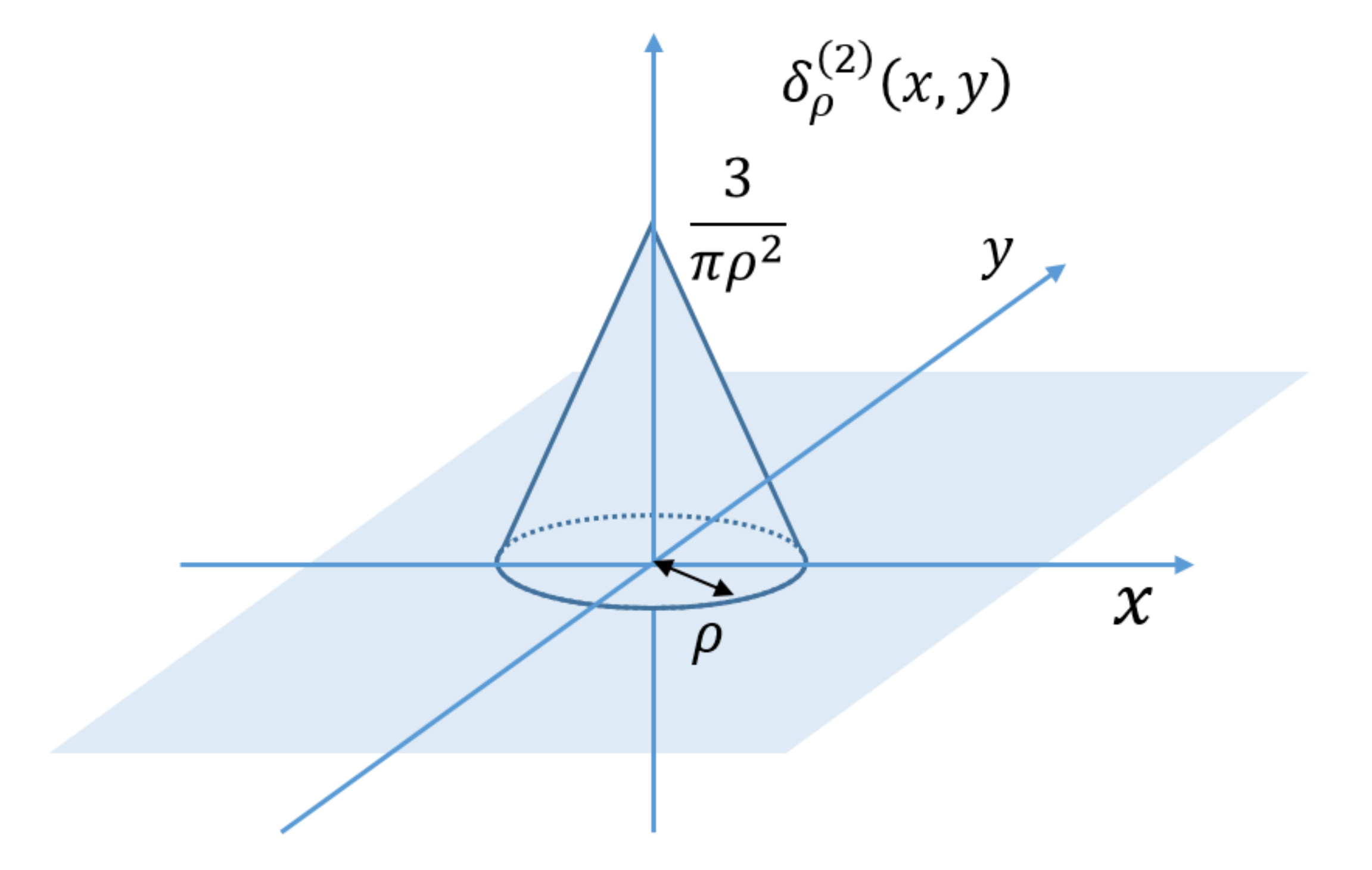}
\end{center}
\caption{A regularization of $\delta^{(2)}(x,y)$}
\label{Reg}
\end{figure}

Here we show the zero mode of the bulk scalar in 1-loop corrected background is localized at the fixed points.
Since the wave function includes the exponential of the delta function, this function is not well defined.
Here we evaluate it by use of an explicit regularization of the delta function.
We regularize the delta function as follows (see Figure \ref{Reg}):

\begin{align}
	\delta_\rho^{(2)}(x,y) = \left\{
	\begin{array}{cc}
		\frac{3}{\pi \rho^2} (1-\sqrt{x^2+y^2}/\rho) & (\sqrt{x^2+y^2} \leq \rho), \\
		0 & (\sqrt{x^2+y^2}>\rho) .
	\end{array}\right.
\label{eq:reg_delta}
\end{align}

We can check $\int dx dy \, \delta_\rho^{(2)}(x,y)=1$ immediately.
\begin{align}
	\int dx dy \,\delta_\rho^{(2)}(x,y) &= \int dr \int d \theta \, r \delta_\rho^{(2)}(r,\theta) \notag \\
	&=\int_0^\rho dr \int d \theta \, r \frac{3}{\pi \rho^2} (1-r/\rho) \notag \\
	&=2\pi \int_0^\rho dr  \, \frac{3}{\pi \rho^2} (r-r^2/\rho) \notag \\
	&= 2\pi \frac{3}{\pi \rho^2} \times \frac{\rho^2}{6} = 1.
\end{align}
The wave function of the zero mode is given by (\ref{even_zero_mode}).
Substituting the regularization \eqref{eq:reg_delta} into the wave function,
we obtain 
\begin{align}
	|\phi_{+,0}(z,\bar{z})|^2 
	\sim
	\begin{cases}
	|f|^2 \prod_{I=1,...,4} |\psi_I(z,\bar z)| \exp\Big\{ \frac{3 k}{\pi \rho^2} (1-|z-z_I|/\rho)  \Big\} & (|z-z_I| \leq \rho)\\
	|f|^2 \prod_{I=1,...,4} |\psi_I(z,\bar z)| & (|z-z_I|> \rho)
	\end{cases}
\end{align}
where $k = \frac{4gq \xi''}{R^2}$ and $\psi_I(z,\bar{z})$ is given by
\begin{align}
|\psi_I(z,\bar{z})|^2\equiv   \Big| \vartheta_1\Big(\frac{z-z_I}{2\pi}\Big| \tau \Big) \Big|^{2gq\xi_I/2\pi} \exp\Big\{ -\frac{gq \xi_I}{4\pi^2 \text{Im}\,\tau}\{\text{Im}\, (z-z_I)\}^2 \Big\} .
\end{align}
We define $\mathcal{D}_I$ as the disc with radius $\rho$ around the fixed points $z_I$.
Since $\psi_I(z,\bar z)$ is finite except for the vicinities of the fixed points,
we can evaluate the norm of the wave function by the sum of integrals on $\mathcal{D}_I$:
\begin{align}
\int_{T^2} dzd\bar z |\phi_{+,0}|^2 = \sum_{I=1,...,4} \int_{\mathcal{D}_I} dzd\bar z |\phi_{+,0}|^2 +  C,
\end{align}
where $C$ is a finite constant, which is almost independent of $\rho$.
(More precisely $\rho$ dependence is sub-leading.)
$C$ is ignorable in the limit of $\rho$ to zero.
In the vicinity of the fixed pints, $\vartheta_1(z - z_I)$ is singular.
It is approximated as
\begin{align}
\vartheta_1\Big(\frac{z - z_I}{2\pi}\Big| \tau \Big) \sim \eta(\tau)^3 (z-z_I),
\end{align}
where $\eta(\tau)$ is the Dedekind eta function.
Thus we can evaluate the wave function around the fixed point $z_I$ by
\begin{align}
|\phi_{+,0}(z,\bar{z})|^2
	\sim
	|f|^2 \left( \prod_{J\neq I} |\psi_J(z_I,\bar{z_I})|^2  \right) 
	\left| \eta(\tau)^3(z-z_I) \right|^{gq\xi_I/ \pi}
	\exp\left[\frac{3k}{\pi \rho^2} (1-|z-z_I|/\rho)\right].
\end{align}
Integral on $\mathcal{D}_I$ is calculated as
\begin{align}
	N_I
	&\equiv
	\int_{0}^{\rho} r dr \int_{0}^{2\pi}d\theta\,
	|\eta(\tau)^3 r|^{gq\xi_I/ \pi} 
	\exp\left[\frac{3k}{\pi \rho^2} (1-r/\rho)\right]
	\nonumber
	\\
	&=
	2\pi |\eta(\tau)|^{3gq\xi_I/ \pi} e^{\frac{3k}{\pi \rho^2}}  
	\left(\frac{\pi \rho^3}{3k} \right)^{2+gq\xi_I/ \pi}
	\int_{0}^{\frac{3k}{\pi \rho^2}} dr'  
	r'^{1+gq\xi_I/ \pi} 
	e^{-r'}
	\nonumber
	\\
	& \sim 
	2\pi |\eta(\tau)|^{3gq\xi_I/ \pi} e^{\frac{3k}{\pi \rho^2}}
	\left(\frac{\pi \rho^3}{3k} \right)^{2+gq\xi_I/ \pi}
	\Gamma\left(2 + \frac{gq\xi_I}{ \pi}\right),
\end{align}
where $\Gamma(z)$ is the gamma function,
and we have approximated the integration range by $\mathbb{R}_+$.
$\Gamma\left(2 + \frac{gq\xi_I}{ \pi}\right)$ is not zero since $gq \xi_I/\pi$ is positive definite.
$N_I$ diverges in the limit of $\rho \rightarrow +0$.
To normalize the zero mode, we obtain
\begin{align}
	f = 
	\left(\sum_{I=1,...,4}N_I
	\prod_{J\neq I} |\psi_J(z_I,\bar{z_I})|^2 \right)^{-1/2} \rightarrow 0.
	\nonumber
\end{align}
Except for $\mathcal{D}_I$, we find $|\phi_{+,0}(z)|^2 = f^2 \prod|\psi_I(z)| \rightarrow 0$ in the limit of $\rho \rightarrow +0$.
Thus the zero mode wave function is localized at the fixed points.
It behaves as a linear combination of the delta functions $\delta(z-z_I)$:

\begin{align}
	|\phi_{+,0}(z,\bar{z})|^2= \sum_{I=1,...,4} C_I \delta^{(2)}(z-z_I).
\end{align}
The coefficients $C_I$ are calculated by the surface integrals of $|\phi_{+,0}(z,\bar{z})|^2$ on small disc $\mathcal{D}_I$: 
\begin{align}
	C_I 
	&= \int_{\mathcal{D}_I} d^2z \,|\phi_{+,0}(z,\bar{z})|^2
	\nonumber\\
	& \sim
	\int_{\mathcal{D}_I}
	|f|^2 \left( \prod_{J\neq I} |\psi_J(z_I,\bar{z_I})|^2 \right) \left|\eta(\tau)^3(z-z_I)\right|^{gq\xi_I/ \pi} 
	\exp\left[\frac{3k}{\pi \rho^2} (1-|z-z_I|/\rho)\right]
	\nonumber\\
	& =
	\frac{N_I \prod_{J\neq I} |\psi_J(z_I,\bar{z_I})|^2 }
	{\sum_{I=1,...,4} N_I
	\prod_{J\neq I} |\psi_J(z_I,\bar{z_I})|^2}.
\label{eq:C_I_reg}
\end{align}
Extracting $\rho$ dependence of $N_I$, we obtain
\begin{align}
N_I \propto e^{\frac{3k}{\pi \rho^2}} \rho^{3(2+\frac{ g q \xi_I}{\pi})},
\end{align}
while $\rho$ dependence of the denominator of \eqref{eq:C_I_reg} is evaluated as
\begin{align}
\sum_{I=1,...,4} N_I \prod_{J\neq I} |\psi_J(z_I,\bar{z_I})|^2 
\sim e^{-\frac{3k}{\pi \rho^2}} \rho^{-3(2+\frac{ g q \xi_{min}}{\pi})},
\end{align}
where $\xi_{min}$ is the minimum of $\xi_1,...,\xi_4$.
If $\xi_I$ is bigger than $\xi_{min}$, $C_I$ vanishes in the limit of $\rho$ to zero.
We obtain $C_I$
\begin{align}
	C_I =
	\begin{cases}
	\frac{\prod_{J\neq I} |\psi_J(z_I,\bar{z_I})|^2 }
	{\left(\sum_{\xi_{I} = \xi_{min}}
	\prod_{J\neq I} |\psi_J(z_I,\bar{z_I})|^2 \right)},~~~~~~(\xi_I = \xi_{min})
	\nonumber\\
	0.~~~~~~~~~~~~~~~~~~~~~~~~~~~~~~~~~~~~(\xi_I > \xi_{min})
	\end{cases}
\end{align}
This is nothing but \eqref{eq:r_I}.
Thus we can evaluate $C_I$ by the absolute value of the wave function near the fixed point.

\section{Modular symmetry of elliptic theta functions }
\label{app:C}

Here, we summarize modular symmetry of elliptic theta functions.
Under the $S$ transformation, they satisfy the relations,
 \begin{eqnarray}
\vartheta_1(0|-1/\tau) &=& -i\sqrt{-i \tau}\vartheta_1(0|\tau) , \qquad 
\vartheta_2(0|-1/\tau) = \sqrt{-i \tau}\vartheta_4(0|\tau) ,  \nonumber \\
\vartheta_3(0|-1/\tau) &=& \sqrt{-i \tau}\vartheta_3(0|\tau) , \qquad 
\vartheta_4(0|-1/\tau) = \sqrt{-i \tau}\vartheta_2(0|\tau) .
\end{eqnarray}
Also, under the $T$ transformation, they satisfy the relations,
 \begin{eqnarray}
\vartheta_1(0|\tau+1) &=& e^{\pi i /4}\vartheta_1(0|\tau) , \qquad 
\vartheta_2(0|\tau+1) = e^{\pi i /4}\vartheta_4(0|\tau) ,  \nonumber \\
\vartheta_3(0|\tau+1) &=& \vartheta_4(0|\tau) , \qquad 
\vartheta_4(0|\tau+1) = \vartheta_3(0|\tau) .
\end{eqnarray}

\end{document}